\documentclass[a4paper]{article}

\usepackage{INTERSPEECH2020}
\usepackage{tabularx,booktabs}

\newcolumntype{s}{>{\hsize=.12\hsize \raggedleft\arraybackslash}X}
\newcolumntype{S}{>{\hsize=.25\hsize \raggedleft\arraybackslash}X} \newcolumntype{B}{>{\hsize=.20\hsize}X}
\newcolumntype{k}{>{\hsize=.04\hsize}X}
\newcolumntype{C}{>{\hsize=.25\hsize \raggedleft\arraybackslash}X}

\title{Multi-view Frequency LSTM: An Efficient Frontend for Automatic Speech Recognition}
\name{Maarten Van Segbroeck, Sri Harish Mallidi, Brian King, I-Fan Chen, Gurpreet Chadha, Roland Maas}
\address{Amazon}
\email{segbrm,mallidih,bbking,ifanchen,gchadha,rmaas@amazon.com}

\begin{document}

\maketitle
\begin{abstract}

Acoustic models in real-time speech recognition systems typically stack multiple unidirectional LSTM layers to process the acoustic frames over time. Performance improvements over vanilla LSTM architectures have been reported by prepending a stack of frequency-LSTM (FLSTM) layers to the time LSTM. These FLSTM layers can learn a more robust input feature to the time LSTM layers by modeling time-frequency correlations in the acoustic input signals. A drawback of FLSTM based architectures however is that they operate at a predefined, and tuned, window size and stride, referred to as `view' in this paper. We present a simple and efficient modification by combining the outputs of multiple FLSTM stacks with different views, into a dimensionality reduced feature representation. The proposed multi-view FLSTM architecture allows to model a wider range of time-frequency correlations compared to an FLSTM model with single view. When trained on 50K hours of English far-field speech data with CTC loss followed by sMBR sequence training, we show that the multi-view FLSTM acoustic model provides relative Word Error Rate (WER) improvements of 3-7\% for different speaker and acoustic environment scenarios over an optimized single FLSTM model, while retaining a similar computational footprint.

\end{abstract}

\section{Introduction}
Thanks to significant advances made in the last decade \cite{hinton2012deep,graves2013speech,xiong2018microsoft,saon2017english,yu2016automatic}, automatic speech recognition (ASR) technology has become an integrated part of our daily lives. Speech recognition has emerged as one of the most natural and intuitive ways to let people control, communicate, and interact with their electronic devices.
The increased adoption of speech technology has led to a surge in demand for speech recognition in voice-driven commercial applications, such as smart home systems, in-car infotainment, personal digital assistants, speech-to-speech translations and voice enabled search systems.
Continuous innovations are required at all levels of the speech recognition engine to ensure that the deployed systems will meet the performance expectations of end-customers in terms of real-time accuracy and latency. 

Recent studies have shown that prepending neural network layers that model spectral correlations and translation invariances in the speech signal, can significantly improve acoustic models trained on large scale, real world speech data. Popular methods are convolutional networks \cite{abdel2014convolutional,wang2017residual}, frequency and time-Frequency LSTMs \cite{li2015lstm,li2016exploring,sainath2016modeling}. These methods learn an advanced feature representation from the acoustic input that is more robust to challenging unseen and noisy scenarios during inference. 

In this paper, we propose to extend the frequency LSTM (FLSTM) approach by proposing a multi-view FLSTM (mvFLSTM) network topology for the acoustic model (AM) or encoder networks of ASR systems. In the mvFLSTM model, multiple stacks of FLSTM layers process the acoustic input signal in parallel, with each stack operating at a different window size and stride along the frequency axis. The outputs of all FLSTM output layers are combined and subsequently projected onto a lower-dimensional affine linear subspace before feeding into the time LSTM layers of the AM. The concept of multi-view or multi-head input layers have been previously exploited in other domains beyond speech recognition \cite{van2013robust,thomas2015improvements,xu2020multi}.

We show the potential of a mvFLSTM for a hybrid ASR system by training acoustic models using the connectionist temporal classification (CTC) \cite{graves2006connectionist} loss followed by sMBR training \cite{vesely2013sequence}. 
Alternatively, the multi-view FLSTM approach presented in this paper, could be applied in a similar fashion to encoder-decoder based sequence-to-sequence models, such as recurrent neural network (RNN) transducer (RNN-T) \cite{graves2012,he2019streaming,li2019improving} and attention-based encoder-decoder architectures \cite{chorowski2015attention,dong2018speech,yeh2019transformer,wang2020reducing}, by prepending the mvFLSTM layers to the encoder. In the scope of this paper, we focus on the hybrid CTC approach.

Extensive experiments were conducted on utterances of real far-field environments by training and evaluating the models on data recorded by voice-controlled far-field devices. With the introduction of a projection layer before the time LSTM layers, the multi-view FLSTM with projection architecture (mvFLSTMp) achieves a superior ASR accuracy over an optimally tuned single FLSTM architecture, without increasing the total number of free parameters in the AM. The latter is an important consideration for streaming devices since reducing the computational requirements of encoding the acoustic model results helps to increase inference efficiency and meet the imposed latency constraints. 

\begin{figure*}[t]
    \centering
    \includegraphics[width=1.65\columnwidth]{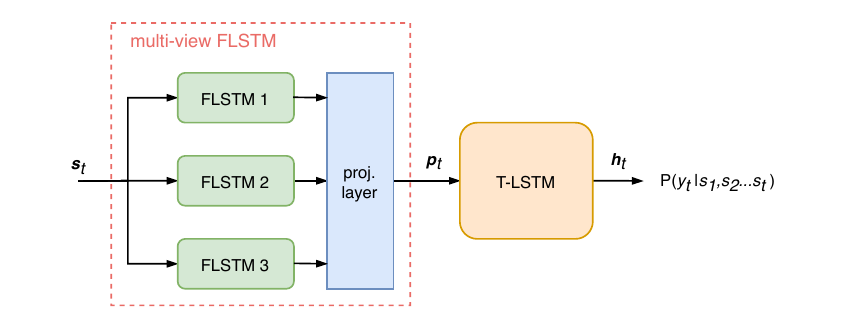}
    \caption{Illustration of an acoustic model architecture for hybrid ASR with multi-view FLSTM and projection layer.}
    \label{fig:mvFLSTM}
    \vspace{-1mm}
\end{figure*}

The paper is outlined as follows. We briefly review FLSTM in Section \ref{section:FLSTM}. Section \ref{section:mvFLSTM} presents the proposed multi-view FLSTM architecture. Experimental evidence is reported in Section \ref{section:experiments} by evaluating on a real-life far-field data. Discussions and conclusions are given in Section \ref{section:discussion} and \ref{section:conclusions}.




\section{Frequency LSTM}
\label{section:FLSTM}
As described in \cite{li2015lstm,sainath2016modeling}, an FLSTM cell has the same mathematical formulation as a time LSTM cell with the difference of sequentially processing the input along the frequency axis. For FLSTM, the sequences are now obtained by windowing the input speech vector $\textbf{s}_t$ with $N$ frequency bins into overlapping chunks with predefined window size and stride. 
In this paper, we will refer to an FLSTM component $i$ as as stack of $K_i$ bidirectional LSTM layers of size $L_i$, processing the input features with a window filter size $F_i$ and stride $S_i$. For each timestamp $t$, unrolling an FLSTM that operates on $N$-dimensioanl input features, results in an output feature $\textbf{v}_{i,t}$ of size
\begin{equation}
	V_i = \frac{N-F_i+S_i}{S_i}*K_i*2
\end{equation}
The multiplication by a factor 2 is required since we are using bidirectional layers.

\section{Multi-view Frequency LSTM}
\label{section:mvFLSTM}
The proposed multi-view FLSTM architecture, as depicted in Figure \ref{fig:mvFLSTM} contains multiple stacks of FLSTM layers, each processing the input features in parallel. By exploiting multiple FLSTM components that apply diverse views along the frequency axis, the mvFLSTM has the potential to capture distinctive spectral patterns in the speech signal. 

Each FLSTM component $i$ learns output features that are complementary to each other and are stacked together into a single feature vector. The stacked output feature is reduced in dimension through a dense layer to project the feature vector onto a lower dimensional space $\textbf{p}_t$, before processing by subsequent encoder layers. After stacking the outputs $\textbf{v}_{i,t}$ of all FLSTM components $i$, we obtain a multi-view output representation $\textbf{v}_{t}$ of size $V=\sum{V_i}$. For practical choices of FLSTM design parameters, $V$ becomes a multiple times larger than the original feature size $N$. Therefore, we add a projection layer to linearly reduce $\textbf{v}_{t}$ to the projection feature vector $\textbf{p}_{t}$ of size $P$. 
The projection layer provides several benefits to the architecture: 
\begin{itemize}
    \item[(i)] when passing $\textbf{v}_{t}$ directly to the encoder's time-LSTM, its large dimensionality causes a significant increase of the total number of trainable parameters in the first layer of the LSTM. For an LSTM input layer of size $M$, with input, output, forget and cell gates, the introduction of a projection layer, with $V*P+P$ weights, reduces the total trainable parameters by $4*M*(V-P)-V*P-P$.
    \item[(ii)] the projection layer removes the dependency of the T-LSTM to FLSTM hidden size, frequency window and stride value. This allows to explore more powerful FLSTM layers with larger layer sizes, and smaller window sizes and strides, that otherwise would not be practical feasible due to memory and computational limits imposed by hardware constraints during training time and inference.
    \item[(iii)] the projection layer acts as a bottleneck layer, to obtain a compact representation of the original input removing undesired redundancy. Some literature studies \cite{grezl2007probabilistic,yu2011improved,yaman2012bottleneck,sainath2012auto} have previously reported that those bottleneck representations could lead to more noise robust representations of speech.  
\end{itemize}
In the next section, we will explore various multi-view FLSTM architectures with up to 3 different views. We will assess the impact of number of views and presence of projection layer on the ASR accuracy and compare the total amount of trainable parameters for different design choices.

\section{Experimental setup}
\label{section:experiments}

\begin{table*}
    \renewcommand{\arraystretch}{1.1}
    \renewcommand{\tabcolsep}{1.5mm}
    \footnotesize
    \centering
    \begin{tabularx}{0.91\textwidth}{kBSSSSSSSC}
        \toprule																	
        {id} & {model}	&	{FLSTM layers}	&	{FLSTM1}	&	{FLSTM2}	&	{FLSTM3}	&	{Proj. dimension}	&	{Total Pars. (M)}	&	{$\Delta$ Pars. (\%)} \\
        \midrule
        01 & LSTM	&	-	&	-	&	-	&	-	&	-	&	 25.6 	& - \\
        \midrule[0.25pt]										
        02 & FLSTM  &   L2x16   &   24/12   &   -   &   -   &   -   &    29.5   &   15.0  \\
        03 & FLSTM	&	L2x16	&	48/24	&	-	&	-	&	-	&	 26.3 	&	2.7	\\
        04 & FLSTM	&	L2x16	&	96/48	&	-	&	-	&	-	&	 24.8 	&	-3.4  \\
        \midrule[0.25pt]									
        05 & mvFLSTM	&	L2x16	&	48/24	&	96/48	&	-	&	-	&	 27.8 	&	8.6 \\
        06 & mvFLSTM	&	L2x16	&	24/12	&	48/24	&	-	&	-	&	 32.5 	&	27.0    \\
        07 & mvFLSTM	&	L2x16	&	24/12	&	96/48	&	-	&	-	&	 31.0 	&	20.8    \\
        08 & mvFLSTM	&	L2x16	&	24/12	&	48/24	&	96/48	&	-	&	 34.0 	&	32.8    \\
        09 & mvFLSTM	&	L2x32	&	24/12	&	48/24	&	96/48	&	-	&	 44.8 	&	75.0    \\
        10 & mvFLSTM	&	L3x32	&	24/12	&	48/24	&	96/48	&	-	&	 44.9 	&	75.3   \\
        \midrule[0.25pt]
        11 & mvFLSTMp	&	L3x32	&	24/12	&	48/24	&	96/48	&	128	&	 24.8 	&	-3.3 \\
        12 & mvFLSTMp	&	L3x32	&	24/12	&	48/24	&	96/48	&	256	&	 26.1 	&	1.7	 \\
        13 & mvFLSTMp	&	L3x32	&	24/12	&	48/24	&	96/48	&	512	&	 28.6 	&	11.7 \\
        \bottomrule																	
    \end{tabularx}
    \vspace{2mm}
    \caption{Various FLSTM and mvFLSTM topologies and their corresponding network size expressed in total number of trainable parameters (Pars.) compared to a baseline time LSTM acoustic model with 5 layers and hidden size of 768. The mvFLSTM topologies contain up to 3 FLSTMs with specified number of layers and layer size (e.g. an FLSTM with 2 FLSTM layers of size 16 hidden cells is denoted as L2x16), and different window size and stride (e.g. 48/24 denotes a window size of 48 frequency bins with a stride of 24 bins). The mvFLSTM models that have an additional projection layer are named mvFLSTMp. }
    \label{tab:network_size}
    \vspace{-4mm}
\end{table*}

The ASR systems are trained on approximately 50K hours of speech data obtained from various voice-controlled far-field devices. The models are trained on low frame rate (LFR) acoustic features \cite{sak2015fast,pundak2016lower} obtained by stacking 3 frames of log-STFT features extracted at a 10~ms frame shift and subsampled every 30~ms. The LFR features are normalized by global Mean and Variance Normalization (MVN) for which the sufficient statistics were derived from the training data.
All models are evaluated on a test set that contains 37,000 utterances. For experimental analyses, the utterances are categorized based on number of speakers per utterance and nativeness. The single-talker (ST) test set contains ~27.5K utterances with only a single speaker per utterance, while the multi-talker (MT) test set has ~9.5K utterances with more than one speaker.  Additional speakers could come from multimedia sources or other people in the same room, for which their speech is not present in the transcripts and hence has to be considered as background noise by the ASR system. Approximately 35.5K utterances are spoken by native speakers, and the remainder ~1.5K utterances are spoken by non-native speakers. We refer to these subsets as the native-talker (NT) and non-native-talker (nNT) test sets. 

Results are reported as relative Word Error Rate Reduction (WERR) compared to a strong hybrid ASR model. The baseline hybrid ASR system has an LSTM-based acoustic model without prepended F-LSTM layers. The HMM states correspond to senones, and use a single-state topology to ensure state that the traversing rate of the LFR features happens at regular 10~ms frame rate. The senone states were tied into 2,608 senones by phonetic decision tree clustering. The AM consists of 5 unidirectional LSTM layers each with 768 cells and a softmax output layer to model the emission probabilities of the 2,608 HMM states. The AM was trained with the Connectionist Temporal Classification (CTC) loss function. A warm start was provided to CTC training by initializing the model weights by pre-training the model with cross-entropy loss (CE) on frame-level senone targets that were obtained by forced alignment. The acoustic models were further improved by sequence-discriminative training using the sMBR criterion as the objective function \cite{vesely2013sequence}. Here, the error signal was computed from the numerator and denominator lattices that were generated for each training utterance by using the corresponding CTC model as the initial seed. The numerator reference state labels are obtained through forced alignment with the reference transcript.

Before processing the LFR features by the FLSTM layers, the LFR input features are permuted such that same frequency bins are grouped together. Experiments not included in the paper, have shown that permuting the features as such results in better modeling performance. Consequently, the FLSTM window sizes $F_i$ were set to be a multiple of 3 such that they contain the same frequency bins for each of the stacked frames. For our experiments, we always chose the window stride $S_i=F_i/2$, which seemed to be a good compromise between FLSTM output size and window overlap. 

\begin{table}[ht]
    \renewcommand{\arraystretch}{1.1}
    \renewcommand{\tabcolsep}{2mm}
    \footnotesize
    \centering
    \begin{tabularx}{0.47\textwidth}{kBsssss}
        \toprule
        id & model &  ST  &   MT  &   NT  &   nNT & Avg.\\
        \midrule
        01  &   LSTM    &   -   &   -   &   -   &   -   &   -   \\
        \midrule[.25pt]
        02  &   FLSTM   &   12.00  &   2.91   &   2.64   &   3.68   &   8.99   \\
        03  &   FLSTM   &   12.95  &   4.86   &   3.85   &   6.27   &   \textbf{10.27}  \\
        04  &   FLSTM   &   10.95  &   3.59   &   2.00   &   5.35   &   8.51   \\
        \midrule[.25pt]
        05  &   mvFLSTM &   14.00  &   4.04   &   4.35   &   7.53   &   10.70  \\
        06  &   mvFLSTM &   13.68  &   5.98   &   4.56   &   10.93  &   11.13  \\
        07  &   mvFLSTM &   12.84  &   6.80   &   4.70   &   6.79   &   10.84  \\
        08  &   mvFLSTM &   12.84  &   8.22   &   4.35   &   11.33  &   11.31  \\
        09  &   mvFLSTM &   13.05  &   7.62   &   4.92   &   10.18  &   11.25  \\
        10  &   mvFLSTM &   14.32  &   8.07   &   4.78   &   8.34   &   \textbf{12.26}  \\
        \midrule[.25pt]
        11  &   mvFLSTMp    &   13.02  &   7.12   &   4.95   &   8.25   &   11.13  \\
        12  &   mvFLSTMp    &   13.47  &   7.21   &   4.85   &   9.15   &   11.45  \\
        13  &   mvFLSTMp    &   15.26  &   7.25   &   5.42   &   8.57   &   \textbf{12.61}  \\
        \bottomrule                                                                                             
    \end{tabularx}
    \vspace{2mm}
    \caption{Results in \% relative WER reduction (WERR) after CTC training for various FLSTM model topologies compared to the baseline LSTM AM (configuration id 01) trained with CTC.}
    \label{tab:CTC}
    \vspace{-4mm}
\end{table}

\section{Discussion}
\label{section:discussion}

Table \ref{tab:network_size} gives an overview of the different FLSTM and multi-view FLSTM layer configurations that we evaluated (with the configuration id specified in first table column). All FLSTM models feed into a time LSTM with the same architecture as that of the baseline AM. The second column denotes the number and hidden size of the FLSTM layers that were used for each stack of FLSTM layers. For the purpose of this paper, each FLSTM stack had the same number of layers and layer size. For the mvFLSTM models, columns 3 to 5 present the window size and stride used in each of the FLSTM components and are denoted as $F_i$/$S_i$. Column 5 indicates whether a projection layer was used and indicates the corresponding dimension after dimensionality reduction. The last two columns respectively show the total amount of trainable parameters for each configuration and its relative increase compared to the baseline LSTM AM. From the table, it is clear that the projection layer in the multi-view FLSTM configurations limits the increase in parameters as discussed above. Models with 3 views and projected dimension up to 512, have similar amount of free parameters as the single-view FLSTM model with window size/stride of 48/24, despite the use of wider and deeper stacks of FLSTM layers (column 2). 
 
The relative reduction in WER on the entire test set and corresponding subsets is shown in Table \ref{tab:CTC}. All WER numbers are compared to the baseline model containing no FLSTM layers. These results suggests that (i) for a single FLSTM configuration, tuning the window size and stride has an important impact on the AM accuracy, (ii) the results improve by combining multiple views of FLSTM components with deeper and wider layers, (iii) the addition of the projection layer does not negatively impact the WER compared to the mvFLSTM models without projection layer, in contrary, we observed a small WER reduction. 

Table \ref{tab:sMBR} shows the recognition results for a selection of model architectures obtained by performing three iterations of sMBR training using the corresponding CTC model as a seed model. From our experiments, we observed that sMBR gave us an additional gains of ~8\% relative WER improvement over the CTC seed models. The multi-view FLSTM architectures, particularly models 09 and 13, outperform the baseline AM and single-view FLSTM model across all subsets. Compared to FLSTM, the mvFLSTM model (id 13) achieved an additional 3\% relative WERR (Avg.) over the best single view FLSTM, and a relative WERR of 7\% on the non-native speakers test set.
Model 13 has a similar multi-view FLSTM configuration as model 09 and a projection layer of size 512. This reduced the total amount of free parameters by 36\% over model 09, but retains the accuracy achieved by the additional front-end modeling of the multi-view FLSTM layers.

\begin{table}
    \renewcommand{\arraystretch}{1.1}
    \renewcommand{\tabcolsep}{2mm}
    \footnotesize
    \centering
    \begin{tabularx}{0.47\textwidth}{kBsssss}
        \toprule
        id & model &  ST  &   MT  &   NT  &   nNT & Avg.\\
        \midrule
        01  &   LSTM    &   15.26  &   7.55   &   4.78   &   9.14   &   12.71  \\
        \midrule[.25pt]
        03  &   FLSTM   &   20.42  &   12.86  &   7.48   &   12.54  &   17.92  \\
        \midrule[.25pt]
        10  &   mvFLSTM &   23.68  &   14.28  &   9.84   &   20.59  &   20.57  \\
        \midrule[.25pt]
        11  &   mvFLSTMp    &   21.68  &   14.50  &   10.69  &   17.08  &   19.31  \\
        12  &   mvFLSTMp    &   22.42  &   13.98  &   10.19  &   14.84  &   19.63  \\
        13  &   mvFLSTMp    &   23.58  &   14.50  &   11.40  &   19.15  &   \textbf{20.58}  \\
        \bottomrule                       
    \end{tabularx}
    \vspace{2mm}
    \caption{Results in \% relative WER reduction (WERR) after CTC+sMBR training for a selection of FLSTM model topologies compared to the baseline LSTM acoustic model trained with CTC.}
    \label{tab:sMBR}
    \vspace{-4mm}
\end{table}

\section{Conclusions}
\label{section:conclusions}
This paper proposed a multi-view frequency LSTM architecture for acoustic modeling in speech recognition, a straightforward extension to the FLSTM topology. In mvFLSTM, the input features are processed in parallel by multiple stacks of FLSTM layers to better capture the various spectral correlations presented in the speech signal. We showed that for a hybrid speech recognition system, an acoustic model with multi-view FLSTM layers that was trained with CTC and sMBR loss on 50K hours of English training data, outperforms the LSTM and FLSTM baseline systems with a relative WERR of 8\% and 3\% respectively, on a evaluation set with more than 37K real-life sentences. Adding a projection layer to the multi-view FLSTM models, significantly reduces the total network size, hence computational load during inference, without any compromise in WER. Future work includes to apply the multi-view FLSTM layers to the encoder networks of encoder-decoder speech recognition models, such as the RNN-T and Transformer encoders.


\bibliographystyle{IEEEtran}

\bibliography{mybib}

\end{document}